\documentclass[aps,prd,superscriptaddress,amsmath,twocolumn,nofootinbib,superscriptaddress,10pt]{revtex4}
\usepackage{color,hhline,psfrag,rotating,amssymb}
\usepackage{dcolumn}
\usepackage{verbatim}
\usepackage{bm}
\begin{document}
\title{The pseudoscalar meson electromagnetic form factor at high $Q^2$ from full lattice QCD}

\author{J.~Koponen}
%\email[]{jonna.koponen@glasgow.ac.uk}
\affiliation{SUPA, School of Physics and Astronomy, University of Glasgow, Glasgow, G12 8QQ, UK}
\affiliation{INFN, Sezione di Tor Vergata, Dipartimento di Fisica, Universit\`{a} di Roma Tor Vergata, Via della Ricerca Scientifica 1, I-00133 Roma, Italy}
\author{A.~C.~Zimermmane-Santos}
\affiliation{SUPA, School of Physics and Astronomy, University of Glasgow, Glasgow, G12 8QQ, UK}
\affiliation{S\~{a}o Carlos Institute of Physics, University of S\~{a}o Paulo, PO Box 369, 13560-970, S\~{a}o Carlos, SP, Brazil}
\author{C.~T.~H.~Davies}
\email[]{christine.davies@glasgow.ac.uk}
\affiliation{SUPA, School of Physics and Astronomy, University of Glasgow, Glasgow, G12 8QQ, UK}
\author{G. P. Lepage}
\affiliation{Laboratory for Elementary-Particle Physics, Cornell University, Ithaca, New York 14853, USA}
\author{A.~T.~Lytle}
\affiliation{SUPA, School of Physics and Astronomy, University of Glasgow, Glasgow, G12 8QQ, UK}
\collaboration{HPQCD collaboration}
\homepage{http://www.physics.gla.ac.uk/HPQCD}
\noaffiliation

\date{\today}

\begin{abstract}
We give an accurate determination of the vector (electromagnetic) form 
factor, $F(Q^2)$, for a light pseudoscalar meson up to squared momentum 
transfer $Q^2$ values of 6 $\mathrm{GeV}^2$ for the first time from 
full lattice QCD, including $u$, $d$, $s$ and $c$ quarks in the sea at 
multiple values of the lattice spacing. 
Our results show good control of lattice 
discretisation and sea quark mass effects. 
We study a pseudoscalar meson made of valence $s$ quarks
but the qualitative picture obtained applies also to the $\pi$ meson, relevant to 
upcoming experiments at Jefferson Lab. 
We find that $Q^2F(Q^2)$ becomes flat in the region 
between $Q^2$ of 2 $\mathrm{GeV}^2$
and 6 $\mathrm{GeV}^2$, with a value well above that of the asymptotic 
perturbative QCD expectation, but well below that of the vector-meson dominance 
pole form appropriate to low $Q^2$ values. 
Our calculations show that we can reach
higher $Q^2$ values in future to shed further light on where the perturbative 
QCD result emerges. 

\end{abstract}

% insert suggested PACS numbers in braces on next line
%\pacs{}
% insert suggested keywords - APS authors don't need to do this
%\keywords{}

%\maketitle must follow title, authors, abstract, \pacs, and \keywords
\maketitle

\section{Introduction}
Hitting one constituent of a bound state with a photon initiates 
a complicated process if the bound state is not to fall apart. 
The momentum gained must be redistributed between all the 
constituents so that the whole convoy can slew round into the new direction. 
A price is paid in terms of a reduced interaction strength
between the photon and the bound state 
and this is known as the electromagnetic form factor - a function of 
the square of the (space-like) 4-momentum, $q^2$, 
transferred from initial to final state. 
When the bound state is a hadron, and held together by the 
strong interaction, the determination of the form factor 
becomes a case study for our understanding of 
Quantum Chromodynamics (QCD) as a function of $q^2$. 
Both experimental measurement~\cite{jlab} and theoretical 
calculation are important. As we show here, lattice QCD,
 now including a realistic QCD vacuum~\cite{ourlatqcd}, 
can provide key 
theoretical results. 

The $\pi$ meson is one of the simplest hadrons, with a valence quark and 
antiquark chosen from $u/d$. 
At small values of squared momentum 
transfer, $Q^2 = -q^2$ up to 0.25 $\mathrm{GeV}^2$, 
its electromagnetic form factor, $F_{\pi}(Q^2)$, 
has been measured directly by scattering from atomic electrons~\cite{amendolia}.  
The form factor can be fitted to a simple pole form in this region and the pole 
mass (close to that of the vector, $\rho$~\cite{Frazer:1959gy}) 
can be related to the r.m.s. electric 
charge radius. Lattice QCD calculations of the $\pi$ form factor
at small values of $Q^2$ 
~\cite{Brommel:2006ww, rbcukqcd, etmpiff, Aoki:2009qn, Nguyen:2011ek, Brandt:2013dua, Aoki:2015pba, Koponen:2015tkr} 
give a theoretical 
determination that agrees well with experiment. 

At the other extreme of the $Q^2$ range, very large values, 
a perturbative QCD treatment of the 
electromagnetic
form factor becomes possible because the process
in which the hard photon scatters 
from the quark or antiquark 
factorises from the distribution amplitudes which 
describe the quark-antiquark configuration in the 
meson~\cite{Lepage:1979zb, Lepage:1980fj}. 
The hard scattering amplitude
is inversely proportional to $Q^2$  
and can be treated perturbatively in QCD because 
a high $Q^2$ photon must be 
accompanied by a high momentum gluon exchange between the meson 
constituents (see Figure 1). 
The asymptotic perturbative QCD prediction, as $Q^2 \rightarrow \infty$, 
is very simple because the distribution amplitude can be normalised using 
the pion decay constant ($f_{\pi}$ = 130.4 MeV)
~\cite{Farrar:1979aw, Efremov:1979qk, Lepage:1979zb}. This gives
\begin{equation}
\label{eq:asympertqcd}
F_{\pi}(Q^2) = \frac{8\pi \alpha_sf^2_{\pi}}{Q^2} 
\end{equation}
but this is not expected to be valid until $Q^2$ values of 
tens of $\mathrm{GeV}^2$ are reached~\cite{Lepage:1980fj}. 
Meanwhile, the approximately constant value of $Q^2F_{\pi}(Q^2)$ from Eq.~(\ref{eq:asympertqcd}) 
is numerically very different 
from the results and trend seen at small $Q^2$. 
This means that there is a large gap to be filled in our understanding, 
extending to relatively high $Q^2$ values~\cite{nsac}. 

For $Q^2$ of a few $\mathrm{GeV}^2$ an indirect experimental method must be 
used to determine $F_{\pi}$ with scattering of electrons 
from the pion cloud around a proton~\cite{Brauel:1977ra, Brauel:1979zk, Ackermann:1977rp}. 
The most recent results from Jefferson 
Lab~\cite{Volmer:2000ek, Tadevosyan:2007yd, Horn:2006tm, Blok:2008jy, Huber:2008id}
have reached $Q^2 = 2.45\, \mathrm{GeV}^2$ but extension 
to $6 \, \mathrm{GeV}^2$ is foreseen~\cite{jlab}, starting 
in 2018, as a key 
experiment (E12-06-101) for the 12 GeV upgrade. 

This is also a $Q^2$ region in which lattice QCD 
can be used to calculate the 
meson electromagnetic form factor directly, as we demonstrate here. 
The method is straightforward, 
and the same for all $Q^2$ values. 
To reach higher $Q^2$ values the participating meson 3-momentum and therefore energy must be 
increased. 
Both statistical errors and systematic errors from discretisation 
effects will then increase, so it is important to have a high statistics 
calculation in a quark formalism with small discretisation errors. 
Previous lattice QCD calculations (see~\cite{Brandt:2013ffb} for a review) 
that include $u$, $d$ and $s$ quarks 
in the sea~\cite{Bonnet:2004fr, Lin:2011sa, Chambers:2017tuf}
have concentrated on having many 
$Q^2$ values for different heavy pion masses 
at one value of the lattice spacing. This has enabled 
studies of pion mass dependence but precluded taking a 
continuum limit. 
See also~\cite{Brommel:2006ww} for a more extensive calculation but 
including only $u$ and $d$ quarks in the sea.  

Here we are able to reach values of $Q^2$ of 6 $\mathrm{GeV}^2$ 
with an accuracy of 10\%
by performing a high statistics calculation at a number of well-separated 
$Q^2$ values. 
Instead of studying $\pi$ 
mesons we work consistently with a `pseudopion', a 
pseudoscalar meson made of 
valence $s$ quarks (denoted $\eta_s$), accurately tuned~\cite{Chakraborty:2014aca} 
on full QCD (with $u$, $d$, $s$ and $c$ quarks 
in the sea) ensembles of gluon field configurations at three values of the lattice spacing 
and two values of the sea $u/d$ quark masses. 
We work with $s$ quarks because it is numerically much faster to 
accumulate high statistics for a precise result, little dependence 
on the sea $u/d$ mass is expected and finite-volume effects are 
negligible~\cite{fkpi}. 
We use the Breit frame where the initial and final mesons 
have opposite spatial momenta, $\vec{p}_i = -\vec{p}_f$ and $Q^2$ is 
maximized for a given $\vec{p}$. By working at values of the lattice 
spacing that range over a factor of 1.7 we are able to show that discretisation 
errors are small for our formalism, even at relatively large $Q^2$, and to extrapolate 
to the zero lattice spacing continuum limit. 

Our $\eta_s$ mesons are qualitatively 
very similar to $\pi$ mesons for the purposes of this study, because the $s$ 
quark is light compared to QCD scales. 
Both the small-$Q^2$ pole form and very high $Q^2$ perturbative 
QCD results for the form factor can be readily determined and thus 
our lattice QCD results provide a clear comparison to these two pictures 
in the region of $0 < Q^2 < 6 \,\mathrm{GeV}^2$. 
In future we can extend this work to even higher $Q^2$ and also calculate other form factors, 
inaccessible to experiment, which can be compared to perturbative 
QCD to understand the $Q^2$ range in which it becomes valid.  
Most importantly, our results show the way to accurate predictions 
for $F_{\pi}$ from 
lattice QCD for the upcoming 
Jefferson Lab experiments~\cite{jlab}. 

\section{Lattice QCD Calculation}
\label{sec:lattice}

The electromagnetic, or vector, form factor for a pseudoscalar meson, $P$, is determined 
from 
\begin{equation}
\label{eq:ff}
 \langle P(p_f) | V_{\mu} | P(p_i) \rangle = F_P(p_i+p_f)_{\mu},
\end{equation}
where $V_{\mu}$ is a vector current coupling to the photon. 
Here we use the temporal component of $V$ and $\vec{p}_i = -\vec{p}_f$ 
so that the right-hand side of eq.~(\ref{eq:ff}) becomes 
$2EF_P$ with $Q^2 = |2\vec{p}_i|^2$. 

The matrix element is determined in lattice QCD by combining information from 
meson `2-point' and `3-point' functions~\cite{DeGrand:2006zz}. 
2-point functions 
tie together quark and antiquark propagators for a correlation 
function that creates a hadron at time
0 and destroys it at time $t^{\prime}$. 
3-point functions combine 3 propagators so that
a meson is created at time 0, its quark 
(or antiquark) carrying momentum $\vec{p}_i$ 
interacts with a photon at time 
$t$ and is scattered into $\vec{p}_f$, with the meson 
being destroyed at time $T$\footnote{
Charge-conjugation symmetry
means that quark-line disconnected diagrams 
vanish in this case~\cite{Draper:1988bp}.}. 
We fit the $t^{\prime}$-, $t$- and $T$-dependence of 
the 2- and 3-point results (averaged over the 
gluon field configurations in an ensemble and including 
all results above a $t_{\mathrm{min}}$ of 3)  
simultaneously to a multi-exponential form
in Euclidean time that includes the set of possible mesons 
made from this valence quark and antiquark~\cite{Koponen:2015tkr}. 
This enables us to isolate the matrix element 
for the ground-state, lightest, meson and relate it 
to the required form factor, whilst making sure that systematic 
effects from the presence of higher mass states in the correlator 
are taken into account. We can normalise the 
form factor by the electric charge conservation requirement that $F_P(Q^2=0)=1$.  

\begin{figure}
\centering
\includegraphics[width=0.48\textwidth]{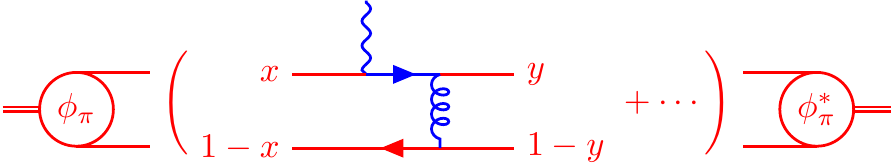}
\caption{
The perturbative QCD description of the $\pi$ electromagnetic form 
factor. $\phi_{\pi}$ represents the distribution amplitude and 
the blue lines indicate the route of high momentum transfer through 
the hard scattering process. 
}
\label{fig:Fpi}
\end{figure}

We use the Highly Improved Staggered Quark (HISQ) 
formalism designed~\cite{HISQ_PRD}, and 
shown~\cite{HISQ_PRL, Dsdecayconst, Donald:2012ga, fkpi}, to have 
very small discretisation errors from the lattice spacing. 
We work on gluon field
configurations generated by the MILC 
collaboration~\cite{Bazavov:2010ru, Bazavov:2012uw}
that include HISQ $u$, $d$, $s$ and $c$ quarks in the sea 
and also have a highly-improved gluon action~\cite{Hart:2008sq}. 
On these configurations we study the $\eta_s$. 
In lattice QCD we can prevent this 
particle from mixing with other isospin zero mesons 
and then its properties can be well determined~\cite{fkpi} and 
it behaves as a pseudopion; its mass 
is 688.5(2.2) MeV and decay constant 181.14(55) MeV. 
Here we determine its vector form factor as 
a function of $Q^2$. 

Table~\ref{tab:params} gives the parameters 
of the gluon field configurations we use, with lattice 
spacing varying from 0.15 fm to 0.09 fm and $u/d$ quark mass 
either twice or five times the physical value, corresponding 
to $M_{\pi} \approx$ 216 or 304 MeV. 
We tune the 
valence $s$ quark mass on each ensemble to obtain the correct 
$\eta_s$ mass~\cite{Chakraborty:2014aca}. We calculate 
$\eta_s$ 2-point functions with a range of spatial momenta  
with magnitude in lattice units up to 0.62, given in Table~\ref{tab:params}. 
These are implemented by using 
the `twisted boundary condition' method~\cite{twist} and are chosen 
to be in the (1,1,1) direction to minimise discretisation effects. 
We use $\eta_s$ mesons made with 
the local $\gamma_5$ (Goldstone) operator; in staggered quark parlance this 
corresponds to spin-taste $\gamma_5 \otimes \gamma_5$~\cite{HISQ_PRD}. 
For our 3-point correlation functions we use a 1-link temporal 
vector current with spin-taste $\gamma_0 \otimes 1$. 

We fit 2- and 3-point 
correlators simultaneously using Bayesian methods~\cite{gplbayes}
to constrain fit parameters and determining the covariance 
between results at different $Q^2$ values. 
The fit forms are~\cite{Donald:2012ga, fkpi} 
\begin{eqnarray}
C_{2pt}(\vec{p}) &=& {\sum}_{i} b_{i}^2(p) \mathrm{f}(E_{i}(p),t^{\prime}) + \mathrm{o.p.t.} \nonumber \\
C_{3pt}(\vec{p},-\vec{p}) &=& {\sum}_{i,j} \big [b_{i}(p) \mathrm{f}(E_{i}(p),t)J_{ij}(Q^2)b_{j}(p) \times \nonumber \\
&& \hspace{4em}\mathrm{f}(E_j(p),T-t)\big ]  
 + \mathrm{o.p.t.} \nonumber \\  
\mathrm{f}(E,t) &=& e^{-Et} + e^{-E(L_t-t)} 
\label{eq:2ptfit}
\end{eqnarray}
The HISQ action gives opposite parity terms (o.p.t.) 
for $\eta_s$ mesons at non-zero momentum; they are similar to 
the terms given explicitly above 
but with factors of $(-1)^{t^{\prime}/a}$. The fit parameters are chosen to be 
the log of the ground-state energy, $E_0$, and the log of 
energy differences between the (ordered) excitations, $i$. 
For our 
kinematic set-up $F_{\eta_s}(Q^2) = J_{00}(Q^2)/J_{00}(0)$, with $J_{00}$ the
ground-state to ground-state amplitude. The division 
by $J_{00}(0)$ provides the normalisation of the lattice current. 
Results for the renormalisation factors inferred from 
$J_{00}(0)$ are given in Appendix~\ref{appendix:renorm}. 

We use priors of 800 $\pm$ 400 MeV for the energy splitting between 
successive excitations and prior widths on amplitudes $b_i$ and $J_{ij}$ of 
 at least 2 times the ground-state value. 
We take results from fits 
that include 6 exponentials where ground-state values and their uncertainties 
have stabilised and we have checked that the prior widths have 
only a minor impact on these uncertainties. Although we are only
interested in ground-state properties here,
our correlators are precise enough to resolve the first excited state. We 
have checked that its mass (around 950 MeV above the ground-state) is in 
reasonable agreement with that for an excited $0^-$ $s\overline{s}$ state 
seen in~\cite{Dudek:2011tt}. Note that we do not expect multi-meson (for example two 
kaon) energy levels to appear in our spectrum since the overlap of such states 
with our single meson operators is very small, 
being suppressed by the volume~\cite{Dudek:2010wm}.  

\begin{figure}
\centering
\includegraphics[width=0.48\textwidth]{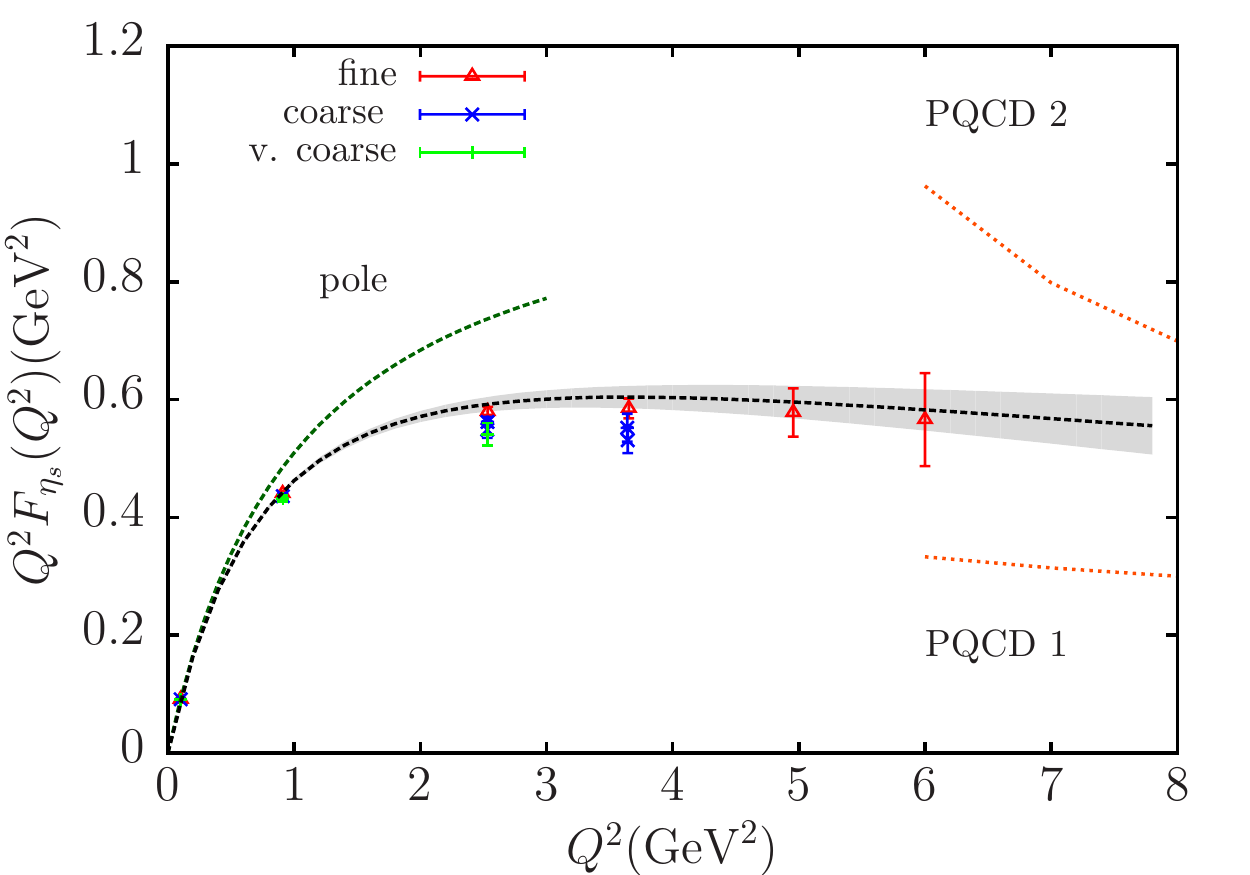}
\caption{
Lattice QCD results for the vector form factor of the $\eta_s$ meson, 
multiplied by $Q^2$ to focus on the large $Q^2$ behaviour, plotted 
as a function of $Q^2$. From coarse to fine: set 1 results are 
given by green pluses, set 2 by blue crosses, 
set 3 by blue bursts and set 4 by red triangles. 
Error bars include
statistical/fit errors and uncertainties from the lattice spacing correlated
between points. The black dashed line and grey band (for $\pm 1 \sigma$) give 
the physical-point curve discussed in the text. 
The green dashed line marked `pole' gives the pole form ($P_{\phi}^{-1}$), 
for comparison. 
The orange dotted line marked `PQCD 1' gives the asymptotic perturbative 
QCD prediction and that marked `PQCD 2' includes non-asymptotic corrections to 
the distribution amplitude discussed in the text.  
}
\label{fig:fq2}
\end{figure}

Results for the (ground-state) form factor
are given in Table~\ref{tab:results} and $Q^2F(Q^2)$ is plotted 
in Figure~\ref{fig:fq2}. Results on different ensembles lie
close to each other, showing that effects from discretisation and different 
$u/d$ masses are very small. Tests of discretisation effects
from studies of the meson energy and decay amplitudes 
as a function of spatial momentum 
are reported in Appendix~\ref{appendix:disc}. 
We also show in Appendix~\ref{appendix:disc} (see Figure~\ref{fig:err}) how
statistical errors in the form factor grow as a function of $Q^2$ and $(Qa)^2$.   
It is in fact the statistical 
errors that provide a practical 
limit to how high in $Q^2$ we can reach here for different values of 
the lattice spacing. 
Note that the finer lattices have 
larger reach in $Q^2$ than the coarse. 

\begin{table*}
\caption{
We use MILC gluon field configurations
~\cite{Bazavov:2010ru, Bazavov:2012uw}, with
$\beta=10/g^2$ the QCD coupling and $L_s$ and $L_t$ the lattice 
dimensions. $w_0/a$~\cite{fkpi}
gives the lattice spacing, $a$, in terms of the Wilson flow parameter,
$w_0$~\cite{Borsanyi:2012zs};
$w_0$=0.1715(9)\,fm from $f_{\pi}$~\cite{fkpi}. 
Set 1 is `very coarse', sets 2 and 3, `coarse' and set 4, `fine'.
$am_l, am_s$ and $am_c$ are the sea quark masses ($m_l\equiv m_u=m_d$)
in lattice units. $am_s^\mathrm{val}$ is the valence $s$ mass and 
$aM_{\eta_s}$ the corresponding $\eta_s$ mass in lattice 
units. 
$n_{\mathrm{cfg}}$ gives the number of
configurations;
16 random-wall time sources on each give high statistics. 
$ap$ gives the magnitude of the meson spatial momentum for 
the form factor at non-zero $Q^2$. 
We further reduce uncertainties  
on set 2 at $pa=0.6$ by averaging over 4 directions. 
We use 3 values of $T/a$ for our 3pt-functions: 
9, 12, 15 on set 1; 12, 15, 18 on sets 2 and 3 and 15,18 and 21 on set 4.
}
\label{tab:params}
\begin{ruledtabular}
\begin{tabular}{ccllllllccr}
Set & $\beta$ &
\multicolumn{1}{c}{$w_0/a$} &
\multicolumn{1}{c}{$am_l$} &
\multicolumn{1}{c}{$am_s$} &
\multicolumn{1}{c}{$am_c$} &
\multicolumn{1}{c}{$am_s^\mathrm{val}$} &
\multicolumn{1}{c}{$aM_{\eta_s}$} &
\multicolumn{1}{c}{$ap$} &
{$L_s/a\times L_t/a$} &
\multicolumn{1}{c}{$n_{\mathrm{cfg}}$} \\
\hline
1 & 5.8 & 1.1119(10) & 0.0130 & 0.0650 & 0.838 & 0.0705 & 0.54028(15) & 0.1243,0.3730,0.6217 & $16\times48$ & 1020 \\
2 & 6.0 & 1.3826(11) & 0.0102 & 0.0509 & 0.635 & 0.0541 & 0.43135(9) & 0.1,0.3,0.5,0.6($\times$ 4 dirns) & $24\times64$ & 1053\\
3 & 6.0 & 1.4029(9) & 0.00507 & 0.0507 & 0.628 & 0.0533 & 0.42636(6) & 0.493,0.591 & $32\times64$ & 1000\\
4 & 6.3 & 1.9006(20) & 0.0074 & 0.037 & 0.44 & 0.0376 & 0.31389(7) & 0.0728,0.218,0.364,0.437,0.509,0.56 & $32\times96$ & 1008\\
\end{tabular}
\end{ruledtabular}
\end{table*}

\begin{table*}
\caption{
Results for the vector form factor, with statistical error, at values of $Q^2$ given in $\mathrm{GeV}^2$.  
}
\label{tab:results}
\begin{ruledtabular}
\begin{tabular}{lllllllllllll}
Set & $Q^2$ & $F(Q^2)$ & $Q^2$ & $F(Q^2)$ & $Q^2$ & $F(Q^2)$ & $Q^2$ & $F(Q^2)$ & $Q^2$ & $F(Q^2)$ & $Q^2$ & $F(Q^2)$ \\
\hline
1 & 0.1012 & 0.9003(9) &  0.9109 & 0.4747(18) & 2.531 & 0.2138(70) & & & & & & \\
2 & 0.1012 & 0.9009(5)   &  0.9111 & 0.4786(10)   &  2.531 & 0.2170(51)    &  3.644 & 0.1456(59) & & & & \\
3 &  & & & & 2.533 & 0.2219(23)  & 3.640 & 0.1517(65)  & & & & \\
4 & 0.1014  & 0.9014(6)  &  0.9091 &  0.4843(9)  & 2.535 & 0.2286(22)  & 3.653 & 0.1602(42)   &  4.956 & 0.1167(82)   & 5.999 &  0.094(13)  \\
\end{tabular}
\end{ruledtabular}
\end{table*}

To determine the form factor in the physical continuum limit we 
must extrapolate in the lattice spacing and sea $u/d$ quark mass. We do 
this using a model-independent parameterisation of the form 
factor now standard in both theory and experiment for semileptonic weak 
decays (see~\cite{Hill:2006ub} for a recent review), mapping the 
domain of analyticity in $t=q^2$ onto the unit circle in $z$. 
Since $z < 1$ we can then perform a power series 
expansion in $z$. 
We take~\cite{Lee:2015jqa} 
\begin{equation}
z(t,t_\mathrm{cut})=\frac{\sqrt{t_\mathrm{cut}-t}-\sqrt{t_\mathrm{cut}}}{\sqrt{t_\mathrm{cut}-t}+\sqrt{t_\mathrm{cut}}}
\end{equation}
where $t_\mathrm{cut}$ in our case is equal to $4M_K^2$. We choose the point 
that maps to $z=0$ to be $q^2=0$, for simplicity; this gives $z_{\mathrm{max}}$ of 0.46 
at $Q^2 = 6 \,\mathrm{GeV}^2$, well below 1. 
Rather than $F(Q^2)$ we work with $P_{\phi}(Q^2)F(Q^2)$, using 
$P_{\phi}(Q^2)=(1 + Q^2/M_{\phi}^2)$. The product $P_{\phi}F$ has reduced $z$-dependence
because $P^{-1}_{\phi}$ is a good match to the form 
factor at small $Q^2$ (the $\phi$ being the $s\overline{s}$ vector 
meson) and it has the correct $Q^{-2}$ dependence at large $Q^2$ 
(but the wrong value: see Figure~\ref{fig:fq2}). 
To combine a $z$-expansion with lattice QCD results we simply allow 
the coefficients in the expansion to have independent $a$- and $m_{\mathrm{sea}}$-dependence.
Adapting the method from~\cite{Koponen:2013tua}, we use the fit function 
\begin{eqnarray}
\label{eq:fit}
&&\hspace{-3em}P_{\phi}F(z,a,m_{\mathrm{sea}}) = 1 + \\
&&\sum_{i=1}^{i_\mathrm{max}} z^i A_i \left[1 + B_i(a\Lambda)^2 + C_i (a\Lambda)^4 + D_i \frac{\delta m}{10} \right] .\nonumber
\end{eqnarray}
Note that the lattice `data' on the left-hand-side include correlations
between results. 
The coefficients $B_i$ and $C_i$ account for dependence on the lattice spacing; 
we take $\Lambda = $ 1 GeV $\approx \sqrt{t_{\mathrm{cut}}}$ to allow 
for $(pa)^2$ and $(pa)^4$ terms in $F$. Independent coefficients at 
each order $i$ allow for $Q^2$-dependent discretisation effects. 
Only even powers of $a$ appear 
in the HISQ formalism and, because we work in the Breit frame with a fixed 
direction for $\vec{p}$, there is only one scale, $p$, that can appear 
coupled with $a$.
Note that, by definition, there are 
no $z$-independent discretisation errors. 
In Figure~\ref{fig:f-asq} of Appendix~\ref{appendix:disc} we 
show results for $Q^2F_{\eta_s}(Q^2)$ at two values of $Q^2$ plotted 
against the square of the lattice spacing, showing more explicitly 
the size of discretisation effects. We also show there 
how well the fit function of eq.~(\ref{eq:fit}) is able to 
reproduce the discretisation effects, including their $Q^2$ dependence. 
$D_i$ accounts for the heavier-than-physical quark masses in 
the sea, using $\delta m = \sum_{u,d,s} (m_q - m_q^{\mathrm{tuned}})/m_s^{\mathrm{tuned}}$
~\cite{Chakraborty:2014aca} and dividing by a factor of 10 to convert this to a suitable 
chiral perturbation theory expansion parameter. 
We take priors on the $B_i$, $C_i$ and $D_i$ of 0.0(1.0) but on $B_1$ of 0.0(5), 
because leading $a^2$ errors are 
suppressed by $\alpha_s$ in the HISQ formalism~\cite{HISQ_PRD}. For the $A_i$, 
the coefficients of the $z$-expansion in the continuum and chiral limits, we take priors of 0.0(2.0), 
twice as conservative as the Bayesian probability function would suggest.  
We use $i_{\mathrm{max}} = 4$; adding higher terms has no impact and neither does 
adding $(a\Lambda)^6$ terms. 

Our fit has a $\chi^2/\mathrm{dof}$ of 0.3. 
The result at $a=0$ and physical quark masses (i.e. $1+\sum A_iz^i$) is plotted (converted back 
to $Q^2$ space) in Figure~\ref{fig:fq2} 
and shows little deviation from the results on the fine lattices.  
The fitted parameters $A_i$ and their covariance matrix are given in 
Appendix~\ref{appendix:params}. 

\section{Discussion/Conclusions}
\label{sec:conclusions}

Figure~\ref{fig:fq2} shows the physical curve for $Q^2F_{\eta_s}(Q^2)$ determined 
from our results for $0 < Q^2 < 7\, \mathrm{GeV}^2$. 
At small $Q^2$ it is compared to the pole form, $P_{\phi}^{-1}(Q^2)$.
Our results show that the physical curve peels away from the pole form 
at $Q^2 \approx 1 \,\mathrm{GeV}^2$ and then lies significantly below it. 
Also plotted in Figure~\ref{fig:fq2}, for $Q^2 > 6 \,\mathrm{GeV}^2$, is the 
asymptotic perturbative QCD form (labelled PQCD 1) of Eq.~(\ref{eq:asympertqcd}), using $f_{\eta_s}$ 
instead of $f_{\pi}$. For $\alpha_s$ we have used $\alpha_s(\overline{\mathrm{\small{MS}}},n_f=3)$
at a scale of $Q/2$, 
since this is the momentum carried by the gluon when the quark and antiquark
share the meson momentum equally. Our physical curve then lies significantly above 
this result 
at $Q^2$ of 6 $\mathrm{GeV}^2$. We expect this qualitative picture of the physical 
curve to be true for 
the pion form factor to be determined in Jefferson Lab experiment E12-06-101 
(the peeling away from the pole form is already apparent~\cite{Huber:2008id}).    

For non-asymptotic $Q^2$ the leading perturbative QCD prediction is modified 
to~\cite{Lepage:1979zb, Lepage:1980fj}:
\begin{equation}
\label{eq:pertqcd}
F_{P}(Q^2) = \frac{8\pi f^2_{P}\alpha_s(Q/2)}{Q^2}\left|1 + \sum_{n=2}^{\infty}a_n^{P}(Q/2)\right|^2. 
\end{equation}
where the sum is over even $n$ for a `symmetric' meson (the $\eta_s$ used here or 
$\pi$ in the isospin limit). 
The $a_n^{P}$, coefficients of an expansion in Gegenbauer polynomials, 
evolve logarithmically to zero as $Q^2 \rightarrow \infty$. 

Lattice QCD calculations have been used to determine $a_2^{\pi}$
~\cite{Arthur:2010xf, Braun:2015axa}
and this changes 
the asymptotic prediction substantially in 
the region of $Q^2$ around 10 $\mathrm{GeV}^2$. 
The calculation of yet higher order corrections is complicated by
operator mixing~\cite{Braun:2007wv}. 
It is important to understand the limitations of the perturbative 
QCD approach here, because
the pion distribution amplitude inferred from $F_{\pi}(Q^2)$ is used in 
other calculations. They appear, for example, in  
light-cone sum rule calculations of the form factor at low $q^2$ 
for the exclusive weak
decay $B \rightarrow \pi \ell \nu$ to determine 
$V_{ub}$~\cite{Khodjamirian:2011ub, Bharucha:2012wy}. 

Figure~\ref{fig:fq2} shows a curve (labelled PQCD 2) that 
uses a shape for the distribution amplitude $\phi_{\eta_s}=(x(1-x))^{\zeta}$ 
at a scale $Q/2 = 2\mathrm{GeV}$ where $x$ is the light-cone momentum 
fraction and $\zeta = 0.52(6)$ is chosen to agree with lattice 
QCD results for $a_2$ for the $\pi$~\cite{Braun:2015axa} (results 
indicate only weak quark mass-dependence, so this should
be a good approximation). 
PQCD 2 is much higher than PQCD 1 at $Q^2 = 6\,\mathrm{GeV}^2$ 
and shows stronger $Q^2$-dependence. 
To obtain a flatter curve in better agreement with our results would require a 
broader distribution amplitude and a higher scale for $\alpha_s$ for less 
evolution. Such curves have been obtained for the $\pi$ in a recent Dyson-Schwinger 
approach~\cite{Chang:2013nia}, and it would be interesting to see if it
can reproduce our results for the $\eta_s$. For this 
purpose we give the parameters for our continuum curve 
in Appendix~\ref{appendix:params}. 

To extend our results to higher values of $Q^2$ is possible on finer 
lattices where a given value of $ap$ corresponds to a higher $|\vec{p}|$ in 
GeV. A $Q^2$ of 12 $\mathrm{GeV}^2$ should be possible on `superfine' lattices with 
$a =$ 0.06 fm, and even 20 $\mathrm{GeV}^2$ at $a =$ 0.045 fm (`ultrafine').  
Lower statistics calculations have already been done on such 
lattices~\cite{McNeile:2010ji, McNeile:2011ng, McNeile:2012qf}. 
See also~\cite{Bali:2016lva} for new methods to reduce uncertainties 
in calculations at high $Q^2$. 
The scalar form factor at high $Q^2$ will give additional information
since perturbative QCD~\cite{Lepage:1979zb, Lepage:1980fj} 
predicts that this should fall more 
rapidly than $Q^{-2}$.  

Perturbative QCD 
(Eq.~\ref{eq:pertqcd}) predicts approximate scaling of the form factor with the square 
of the decay constant, 
and we can test this in lattice QCD 
as we reduce the pseudoscalar meson mass towards that of the $\pi$. 
This scaling may set in before the $Q^2$-dependence becomes 
clearly that of perturbative QCD. 
{\it If} that is the case we can use our results here, 
rescaling by $(f_\pi/f_{\eta_s})^2$, to predict a value for $Q^2F_{\pi}(Q^2)$ 
in a flat region from $2-6 \,\mathrm{GeV}^2$ of $\approx 0.3 \, \mathrm{GeV}^2$. 

{\it Acknowledgements.} We are grateful to the MILC collaboration
for the use of their
gauge configurations and code and to B. Chakraborty and D. Hamilton
for useful discussions.
Our calculations were done on the Darwin Supercomputer
as part of STFC's DiRAC facility jointly
funded by STFC, BIS
and the Universities of Cambridge and Glasgow.
This work was funded by a CNPq-Brazil scholarship,
the National Science Foundation,
the Royal Society, the Science and Technology Facilities Council 
and the Wolfson Foundation.

\appendix
\section{Renormalisation factors}
\label{appendix:renorm}

\begin{table}
\caption{Results for the renormalisation factor $Z_V$ which multiplies the lattice
temporal 1-link vector 
current used here to normalise the form factor fully nonperturbatively. 
The values are obtained from our fits at $Q^2=0$, using $Z_V = 1/J_{00}$.}
\label{tab:ZV}
\begin{tabular}{ll}
Set & $Z_V^{s\overline{s}}$ \\
\hline 
1 & 1.3892(15)    \\
2 &  1.3218(7)   \\
3 &  1.3179(7)   \\
4 &  1.2516(9)   \\
\end{tabular}
\end{table} 

Table~\ref{tab:ZV} gives the values of the vector current renormalisation factor, $Z_V$, 
for each ensemble inferred from electric charge conservation at $Q^2=0$. 
The vector current we use is a 1-link current in the time direction, made gauge-invariant 
by the includion of an APE-smeared gauge link. 
The values for $Z_V$ show the expected qualitative behaviour, slowly falling towards 1 
on finer lattices. 

\begin{figure}[h]
\centering
\includegraphics[width=0.42\textwidth]{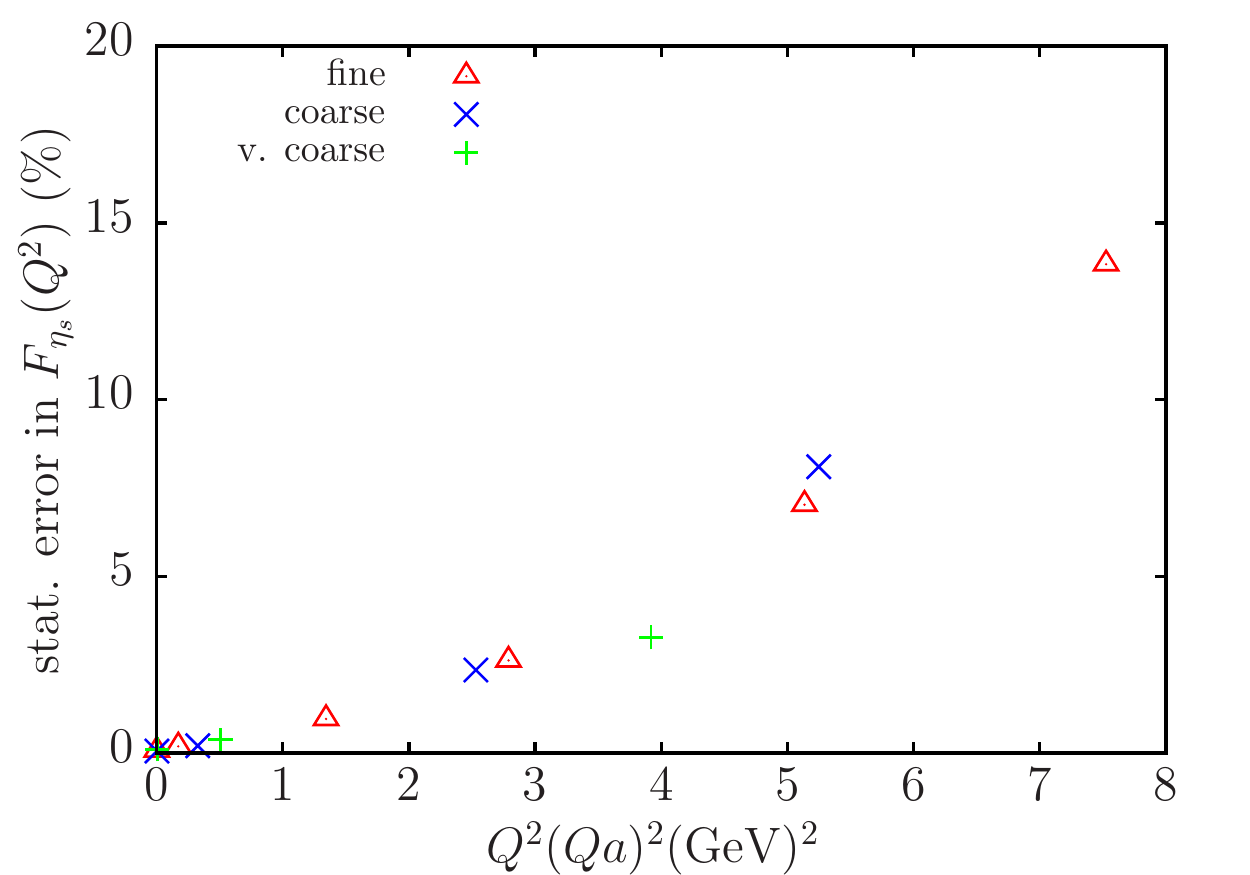}
\caption{
The statistical uncertainty that we obtain in the form 
factor plotted as a function of $Q^2(Qa)^2$ for sets 1, 2 and 4 
(green pluses, blue crosses and red triangles respectively).
For set 2 at $pa=0.6$ we have adjusted the error to be that 
for one spatial momentum direction (instead of 4) to match 
the statistics of the other points. 
Note that these results are specific to the Breit frame and 
teh values of $T$ used here. 
}
\label{fig:err}
\end{figure}

\begin{figure}[h]
\centering
\includegraphics[width=0.42\textwidth]{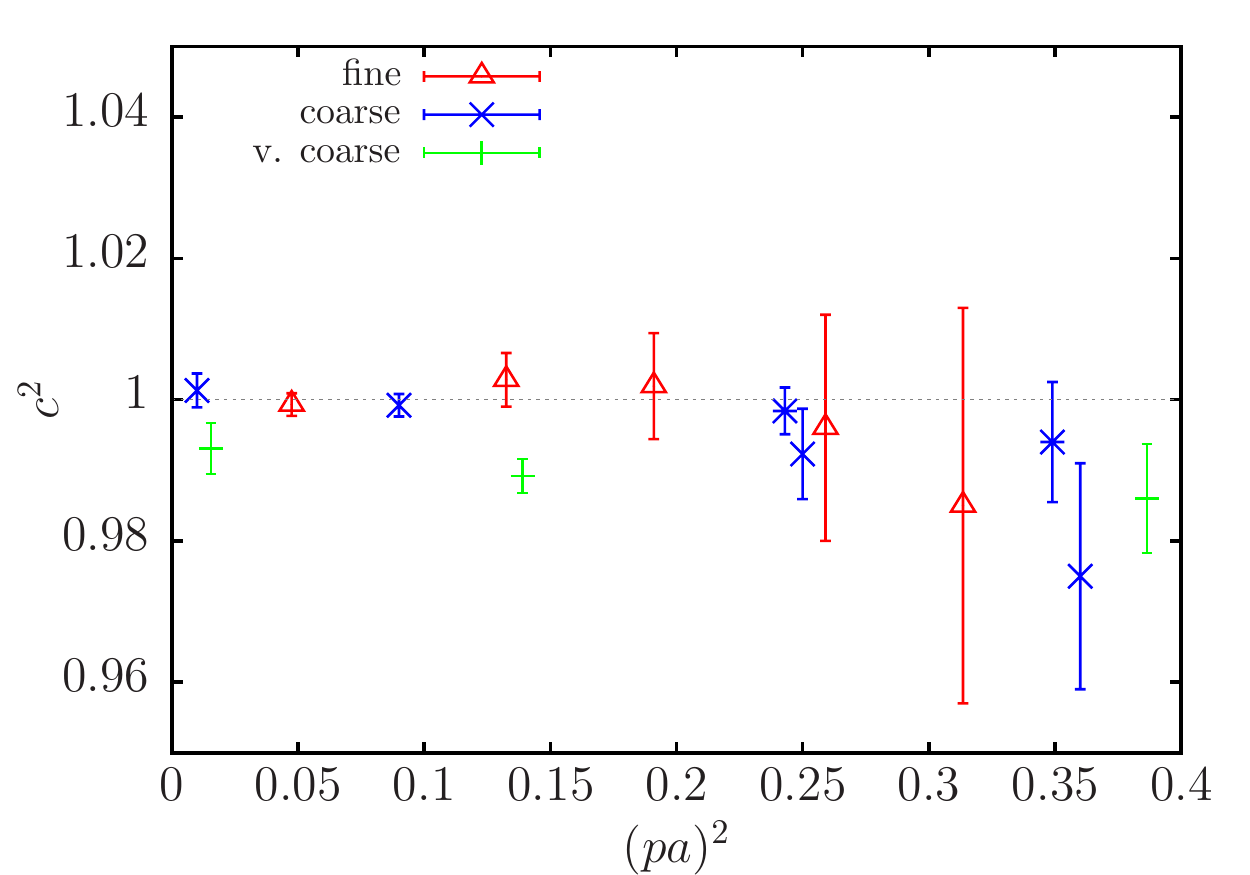}
\caption{
The speed of light, defined in the text, determined from our
$\eta_s$ meson energies as a function of the square of their spatial 
momentum in lattice units. Note the expanded $y$-axis scale. 
Results from very coarse set 1 are given by green pluses, coarse 
set 2 by blue crosses and set 3 by blue bursts, and fine set 4 
by red triangles. 
}
\label{fig:csq}
\end{figure}

\begin{figure}[h]
\centering
\includegraphics[width=0.42\textwidth]{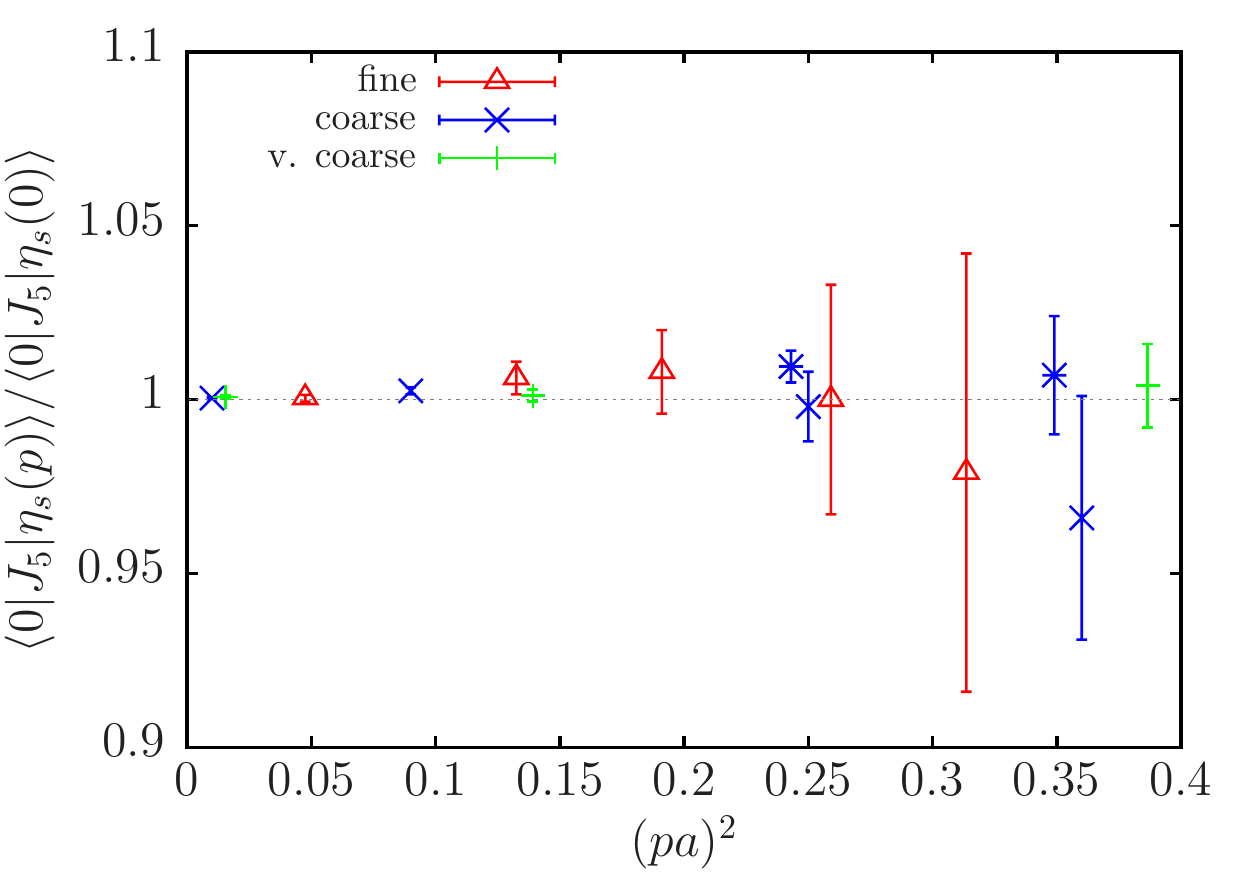}
\caption{
The ratio of the matrix element between vacuum and $\eta_s$ 
of the pseudoscalar density at non-zero spatial momentum to 
that at zero momentum, as a function of the square of 
the meson spatial momentum in lattice units. Symbols as for Fig.~\ref{fig:csq}. 
}
\label{fig:amp}
\end{figure}

\begin{figure}[h]
\centering
\includegraphics[width=0.42\textwidth]{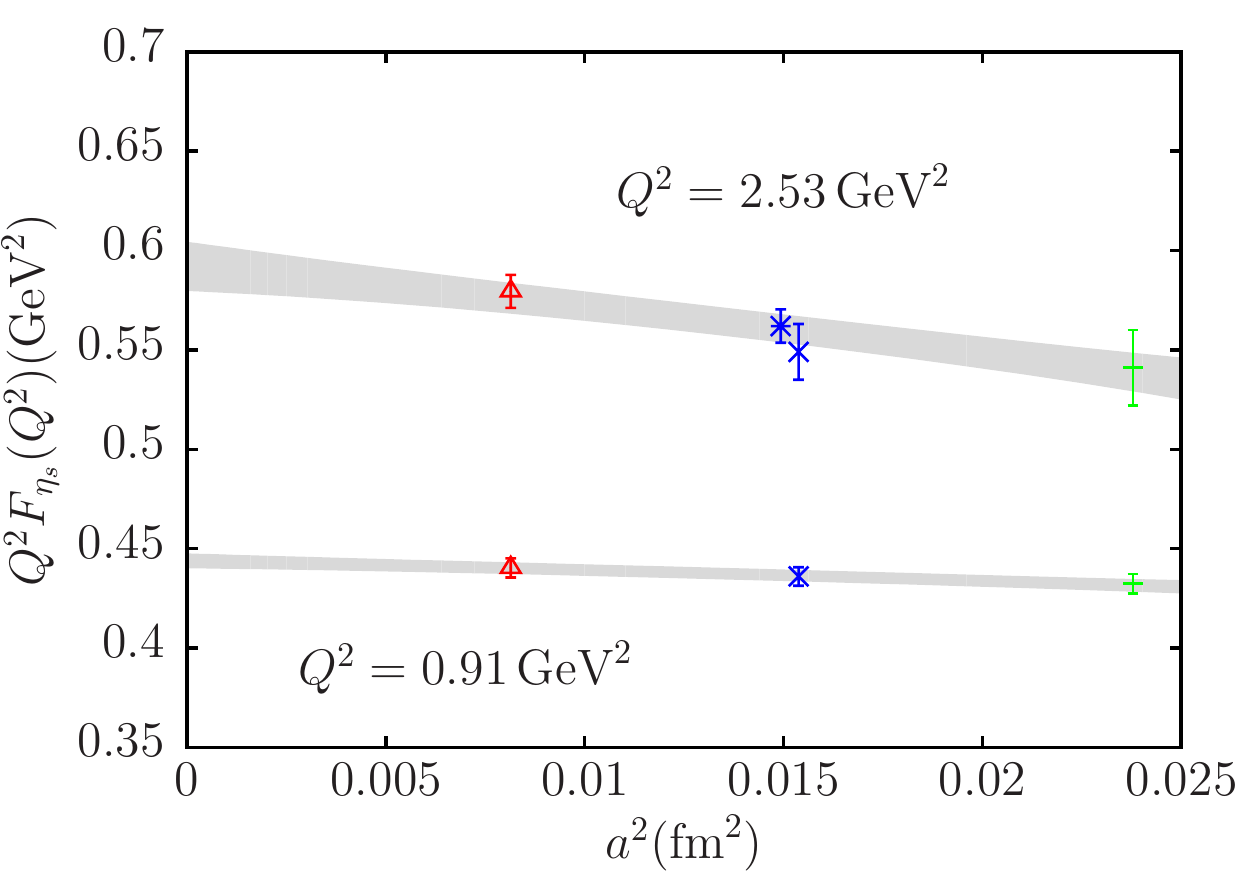}
\caption{
We show here two slices through Figure~\ref{fig:fq2} at two different 
values of $Q^2$, 2.53 $\mathrm{GeV}^2$ and 0.91 $\mathrm{GeV}^2$, 
plotted against the square of the lattice spacing. 
The points (symbols  as for Fig.~\ref{fig:csq}) 
show results at appropriate nearby values of $Q^2$ (see Table~\ref{tab:results}) 
and include correlated uncertainties from the lattice spacing.
The grey bands show our fit result using eq.~(\ref{eq:fit}), now 
as a function of $a^2$ but with $\delta m$ set to zero. 
}
\label{fig:f-asq}
\end{figure}

\section{Tests of statistical errors and discretisation effects}
\label{appendix:disc}

In Figure~\ref{fig:err} we show how the statistical error in the form factor 
result grows with $Q^2$. The results from different lattice spacing values 
(for approximately the same number of configurations, spatial lattice volume and 
smallest $T$ value in physical units)
appear to lie on a universal curve as a function of $Q^2(aQ)^2$. 
The curve is approximately quadratic,
showing that uncertainties degrade rapidly at large $pa$ values. 
However the same $Q$ can 
be reached with smaller $pa$ on finer lattices, moving down the curve. 
The plot helps to
predict the statistical accuracy that will be obtained from calculations 
on other lattices using the same (Breit) frame. 

A good test of discretisation errors is to study the ground-state 
meson energy as a function of spatial momentum and compare the 
speed of light (in units of $c$) obtained from $(E^2-M^2)/\vec{p}^{\, 2}$ 
to the expected value of 1.0 in the absence of systematic discretisation effects. 
The ground-state energy, $E$, is given by $E_0(p)$ and 
the mass, $M$, by $E_0(0)$ from the 2-point fit function of eq.~(\ref{eq:2ptfit}). 
Figure~\ref{fig:csq} shows results from our combined fits to the 
2-point and 3-point $\eta_s$ correlators 
used for our analysis. 
We see that, at the level of
 our statistical uncertainties (at most 3\%), the speed of light shows 
no significant deviation from 1 even  
at the highest momenta that we use for our coarse and fine lattices. For very 
coarse set 1 there is a small (1\%) but significant discrepancy at $pa=0.373$. 
This is consistent with discretisation effects being dominated by $(ma)^4$ terms and 
therefore 3 times larger on the very coarse lattices than on the coarse. 
The statistical uncertainties increase with $(pa)$ 
because the variance of the finite-momentum correlator overlaps with, and so is controlled by, the 
exponential behaviour of the square of the zero-momentum correlator. 
This behaviour is similar to that plotted in Figure~\ref{fig:err} for 
the form factor. 
The uncertainties on the fine lattices at a given value of $(pa)$ are larger than those 
on the coarse lattices, but the values of $(pa)$ correspond to a larger value of 
$|\vec{p}|$, so the accuracy on the finer lattices translates into a 
larger reach in $Q^2$. 
Comparison of the two coarser lattices shows that the larger volume lattices have smaller 
statistical uncertainties at a given $(pa)$ from volume-averaging. 

A further test is to study the ratio of the matrix element of the pseudoscalar density, $J_5$, between 
the vacuum and the $\eta_s$ meson at non-zero spatial momentum to that at zero momentum.
Since the matrix element should be independent of momentum, we expect a result of 1.0. 
The matrix element is determined from the fitted amplitudes denoted by $b_i$ 
in eq.~(\ref{eq:2ptfit}), using 
\begin{equation}
\label{eq:amp}
\frac{\langle 0 | J_5 | \eta_s(p) \rangle}{\langle 0 | J_5 | \eta_s(0) \rangle} = \frac{b_0(p)}{b_0(0)}\sqrt{\frac{E_0(p)}{E_0(0)}}   
\end{equation}
Figure~\ref{fig:amp} shows our results, with a very similar qualitative picture to that of 
Figure~\ref{fig:csq} and again showing excellent control of discretisation effects in the 
HISQ formalism. 

Finally, in Figure~\ref{fig:f-asq}, we illustrate the discretisation errors visible 
in the results for $Q^2F_{\eta_s}(Q^2)$ plotted in Figure~\ref{fig:fq2}. The figure 
shows results at two different values of $Q^2$ for which we have calculations at 
three different values for the lattice spacing (see Table~\ref{tab:results}), 
plotted against the square of the lattice spacing. 
The grey band shows the fit results from eq.~(\ref{eq:fit}) (for $\delta m = 0$) 
at these values of $Q^2$, as a function of $a^2$. We see that discretisation effects, 
although small, are clearly visible. Our fit form has no difficulty in fitting them 
and capturing their $Q^2$-dependence. The accuracy of our results at multiple 
$(Qa)$ and $a$ values is what allows us good control over the continuum limit for the range 
of $Q^2$ values that we cover here.

\section{Parameters of the fit function}
\label{appendix:params}

We give below the values of the fitted parameters, $A_i$ and their covariance 
matrix obtained in the continuum and chiral limit of 
our results. 
In this limit we have (see eq.~\ref{eq:fit}):
\begin{equation}
\label{eq:fitcurve}
P_{\phi} F (z) = 1 + \sum_{i=1}^{4}A_i z^i
\end{equation}
We find :
\begin{eqnarray}
\label{eq:Avals}
A_1 &=& -0.387(59) \nonumber \\
A_2 &=& -0.87(26) \nonumber \\
A_3 &=& 0.4(1.0) \nonumber \\
A_4 &=& -0.5(1.7) 
\end{eqnarray}
Only $A_1$ and $A_2$ are obtained with significance from the fit. 
The $A_i$ have covariance matrix:
\begin{equation}
\left[ \begin{array}{cccc} 0.003472 & -0.008100 & 0.007133 & 0.000999 \\ -0.008100 & 0.068858 & -0.168850 & 0.151820 \\ 0.007133 & -0.168850 & 1.021326 & -1.433623 \\ 0.000999 & 0.151820 & -1.433623 & 2.81513 \end{array} \right] 
\end{equation}

From the fit function for $F$ we can readily derive results 
also for derivatives of $F$ or $Q^2F$. For example the mean square 
electric charge radius is given by 
\begin{eqnarray}
\langle r^2 \rangle_{\eta_s} &=& \left. 6 \frac{dF}{dq^2}\right|_{q^2=0} \\
&=& \frac{6}{M^2_{\phi}}\left( 1 -\frac{A_1}{4}\frac{M^2_{\phi}}{t_{\mathrm{cut}}}\right)
\end{eqnarray}
Thus, from our results, we see that the mean square electric charge radius of 
the $\eta_s$ is 
a factor of 1.103(16) larger than the naive expectation from the $\phi$ 
mass. Translating this into units of fm gives
\begin{equation}
\langle r^2 \rangle_{\eta_s} = 0.248(4) \, \mathrm{fm}^2 . 
\end{equation}
This is, not surprisingly, significantly smaller than the 
mean square electric charge radius of the $\pi$ meson, for which
the Particle Data Group give an average of 0.452(11) $\mathrm{fm}^2$~\cite{pdg}. 

The slope of $Q^2F$ is given from the fit parameters as: 
\begin{eqnarray}
\frac{d(Q^2F_{\eta_s})}{dQ^2} &=& F_{\eta_s} - \frac{Q^2F_{\eta_s}}{P_{\phi}M_{\phi}^2} \\
&+& \frac{Q^2}{P_{\phi}}\frac{(1-z)\sum_i iA_iz^{i-1}}{2\sqrt{(t_{\mathrm{cut}}+Q^2)}(\sqrt{t_{\mathrm{cut}}+Q^2}+\sqrt{t_{\mathrm{cut}}})} \nonumber
\end{eqnarray}
Evaluating this at $Q^2 = 6 \,\mathrm{GeV}^2$ gives  -0.014(8), 
consistent with zero (i.e. a curve for $Q^2F$ that is flat at this point).

\bibliography{hiq2}

\end{document}